\documentclass{sig-alternate}

\usepackage{type1cm}     
\usepackage{graphicx}     
\usepackage{xspace}     
\usepackage{balance}     
\usepackage{booktabs}     
\usepackage[bf,tableposition=top]{caption}     
\usepackage[hyphens]{url}     
\usepackage{bold-extra}     
\usepackage{siunitx}          
\usepackage[vlined,linesnumbered,ruled,noend]{algorithm2e}     
\usepackage[bookmarks, pdftex, colorlinks=false]{hyperref}     
\usepackage{epstopdf}
\usepackage{microtype} 
\usepackage{paralist} 
\usepackage{subcaption} 

\newcommand{\spara}[1]{\smallskip\noindent{\bf{#1}}}

\begin{document}

\title{Quote RTs on Twitter: Usage of the New Feature for Political Discourse\thanks{Accepted at ACM Web Science 2016.}}
\author{
	Kiran Garimella\\
	Aalto University\\
	Helsinki, Finland\\
	kiran.garimella@aalto.fi
	\and
	Ingmar Weber\\
	Qatar Computing Research Institute\\
	Doha, Qatar\\
	iweber@qf.org.qa
	\and
	Munmun De Choudhury\\
    	Georgia Tech\\
	Atlanta, Georgia\\
	munmund@gatech.edu\\
}

\maketitle

\begin{abstract}
Social media platforms provide several social interactional features.
Due to the large scale reach of social media, these interactional features help enable various types of political discourse.
Constructive and diversified discourse is important for sustaining healthy communities and reducing the impact of echo chambers.
In this paper, we empirically examine the role of a newly introduced Twitter feature, `quote retweets' (or `quote RTs') in political discourse, specifically whether it has led to improved, civil, and balanced exchange. Quote RTs allow users to quote the tweet they retweet, while adding a short comment. Our analysis using content, network and crowd labeled data indicates that the feature has increased political discourse and its diffusion, compared to existing features. We discuss the implications of our findings in understanding and reducing online polarization.

\end{abstract}

\section{Introduction}


When it comes to evaluating the impact of the web in general and social media in particular on political discourse at large, there are two broad schools of thought: cyber-optimists and cyber-pessimists. Put simply, cyber-optimists believe that the free flow of ideas online will lead to a more educated populace and stronger democracies. Cyber-pessimists on the other hand believe that digital communication technologies only lead to more polarized opinions by facilitating online ``echo chambers'' and by helping to spread misinformation.

Giving weight to the view held by cyber-pessimists, in recent years there has been a rise of trolls and abuse in social media, particularly in partisan political discourse\footnote{\url{http://www.bbc.com/news/blogs-trending-35111707}}. Especially the comment sections on political news websites often serve as example of ``less than ideal'' political discourse. Only half ironical, this has led to observations such as Godwin's Law\footnote{\url{https://en.wikipedia.org/wiki/Godwin's_law}}, stating that in any online discussion, regardless of topic or scope, someone will sooner or later compare someone else to Hitler. Trolling has even become a tool of international politics with Russia being accused of operating ``Web Brigades''
or troll factories\footnote{
\url{https://en.wikipedia.org/wiki/Web_brigades}}. This frequent abuse of online comment sections has led to many news sites disabling the feature altogether\footnote{\url{http://edition.cnn.com/2014/11/21/tech/web/online-comment-sections/}}. Looking beyond comments, hashtags on Twitter have occasionally also become a source for online abuse. As an example, \#Gamergate was used to defame and outright threat the female game developer Zo\"e Quinn and others\footnote{\url{https://en.wikipedia.org/wiki/Gamergate_controversy}}.

In communication theory, it is well-known that the medium is not separate from the message. In fact, Marshall McLuhan viewed the interplay between what the medium technically allows and the message that it is being used for as so tight that he coined the phrase ``the medium is the message'' \cite{mcluhan1964}. Therefore one might hypothesize that a new feature, i.e., a modification to the medium, should lead to a change to the type of messages that are being transmitted.

Starting from these observations on (i) how features aimed at facilitating online communication can be and have been abused, and (ii) how it is impossible to detangle the medium and the message, we present an analysis of Twitter's newly introduced ``quote retweet'' feature\footnote{\url{https://support.twitter.com/articles/20169873}}.

\begin{figure}
\centering
\includegraphics[width=0.25\textwidth]{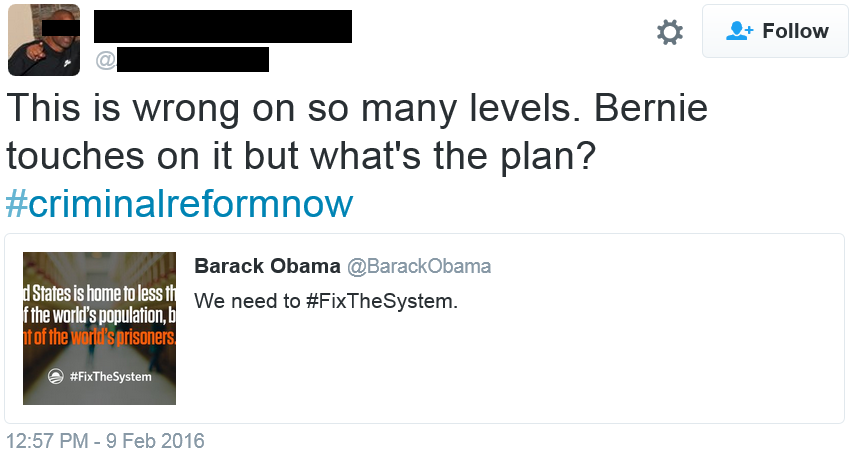}
\caption{Example quote RT with an added comment.}
\label{fig:quotert_example}
\vspace{-2mm}
\vspace{-\baselineskip}
\end{figure}

`Quote Retweets' (or Quote RTs) are a new feature introduced by Twitter in April 2015.
A normal retweet creates a verbatim copy of the original tweet and is typically used as a sign of agreement and endorsement between the retweeting and retweeted users \cite{weberetal13asonam}. 
However, this new feature allows users to quote a tweet while adding their own comment, thereby opening up a number of new use cases. Figure~\ref{fig:quotert_example} shows an example of a Twitter user quoting a tweet by @BarackObama and adding a comment.

A priori, this feature can be used in a number of ways. For example, it can be used to bring the quoted tweet to the attention of a particular user by adding their at-handle in the comment, as they will then be notified of the quoting tweet. This type of usage could lead to a better information propagation in the network. However, it can also be used in less positive ways, Ex., by publicly ridiculing or shaming the quoted person through a negative or even hateful comment. We want to understand how this new feature which facilitates commenting is used and if, overall, it has more of a positive or negative impact on political discourse.





\section{Related Work}


Social media has become an important platform for political discourse. Major political figures have official verified accounts on Twitter and are often in direct contact with their followers; ordinary individuals can take to social media almost anytime, anywhere to voice their opinions, support and critique political events or leanings. Thus, the attributes of social media based political campaigning and political discourse, such as on Twitter, Facebook and Youtube have been examined extensively in prior literature~\cite{cogburn2011networked, robertson2010off, kushin2009getting, halpern2013social}.  

However, despite these affordances, social media platforms have undergone considerable scrutiny in terms of their ability to allow diversified exchange of thoughts and opinions, especially around politically oriented topics~\cite{gilbert2009blogs}. The increasingly involved role of algorithms in social media content personalization as well as varied levels of engagement promoted by different interactional features of the platforms have been argued to be behind the creation of ideological ``echo chambers''~\cite{garrett2009echo}.  Our work is situated in this body of work wherein we examine how the newly introduced `quote RT' feature on Twitter is utilized for political discourse and for sharing of political opinions in the larger social network.

Prior work has also explored the ways and practices that drive the usage of different interactional features of social media platforms for discourse. Retweeting behavior has been the most extensively studied. In an early work, Boyd et al.~\cite{boyd2010tweet} performed a user study to understand why users retweet and found out that users have a wide range of use cases for retweeting, ranging from personal gains (Ex. increasing followers) to spreading information. For a detailed survey on various studies covering retweet behavior, refer to~\cite{metaxas2015retweets}.



However, the usage of the retweet feature on Twitter has been changing ever since the inception of Twitter.
Kooti et al.~\cite{kooti2012emergence} study the emergence of the different forms of retweet behavior and their convergence to a single dominant behavior. This study highlighted the importance of network on the spread and adoption of various information dissemination conventions, and also noted the importance of design changes in Twitter's interactional features in this process. A detailed analysis of various non-conventional ways of retweeting and their impact is discussed in~\cite{azman2012dark}.
Prior to the introduction of quote RTs, users could not comment on a tweet while retweeting. Retweeting almost always signaled endorsement \cite{metaxas2015retweets}. To cope with this, the use of `edited retweets' became prevalent on Twitter. Mustafaraj et al.~\cite{mustafaraj2011edited} studied the prevalence and importance of edited retweets on Twitter and found that edited retweets could be as high as 30\% of the total retweets (in 2011).
%
Unlike retweets where there have been different conventions for the same action in the past (RT), in case of quote RTs we have one convention (in the twitter interface) but different use cases. We present one of the first examinations of the usage of this new feature in political discourse. We also present how these new use cases, which were mostly not possible previously, have enabled new forms of political discourse.


Finally, although interactional features like retweets in general have enabled rich discourse on a variety of topics, they have also been employed for deviant online activities. 
Being an open forum with an ease to conceive personal identity has led to the rise of trolls on different online platforms, Ex., Wikipedia~\cite{shachaf2010beyond}, forum comments~\cite{reader2012free, cheng2015antisocial}, online games~\cite{blackburn2014stfu}, and 
Facebook~\cite{rowe2015civility,halpern2013social}. Building on this line of research, one of the contributions of this paper is characterizing the civility of political discourse on Twitter via the quote RT feature. 

\section{Dataset description}

To define our scope of ``politics'', we used Followerwonk.com and a simple keyword search over Twitter users' bios for political words (Ex., `politics', `democrat', or `republican') to get political users. We limited our search to those users who had at least 100,000 followers as we wanted to limit our scope to well established online personas who are most likely to be addressed in political tweets. 
This resulted in a total of 629 political users. We then manually cleaned the list and restricted the set of users only to US politics, removing news sites, journalists and non-US politicians, resulting in 192 users. We refer to these users as ``seed users''. This set includes all the major political players in the US, including all the candidates in the 2016 presidential race.
Using the full Twitter Firehose, we collected all the tweets by these seed users (254,684), as well as mentions (5.9M), replies (1.2M) and retweets (7M, including quote RTs from April) of these users between Jan--Sep 2015. For the rest of the paper, we only consider only these tweets and users who are involved in these tweets.


\section{Characteristics of Quote RT} 
\label{sec:usage}

In this section, we analyze the various types of use cases that the new quote RT feature has enabled and try to gauge: (i) how these new use cases are different from existing forms of communication and (ii) whether some of these new ways of communication allow an increased political discourse.


As a first step, we manually inspected a random sample of 500 quote retweets from the account @BarackObama and observed the following three major use cases (the choice of this use case was motivated by the observation that these tweets are likely to reflect diverse political discourse):


\spara Opinion - Users expressing opinion in the context of the original tweet. An opinion could also include agreement, disagreement or just stating a fact. Ex.:~\url{https://twitter.com/oren_cass/status/596047822451056641}.

\spara Public Reply - Quote retweets are sometimes replies to the original tweets. Instead of using the `reply' button on the original tweet, this feature is being used to reply and tweet so that all the users followers can see the reply\footnote{Replies only show up in the followers' home timelines if they follow \emph{both} users in the conversation \url{https://support.twitter.com/articles/119138}.}. Ex.: \url{https://twitter.com/AaronDriver5/status/624029124063985664}.

\spara Forwarding - In this case, users use the feature to `forward' particular tweets to other users. This is similar to tagging friends, to include them in a conversation or to get their attention to a particular tweet.
Ex.: \url{https://twitter.com/KillerKaleeb/status/595791426883051520}.



Although these may not be an exhaustive list of ways in which people have been using this feature, from our manual analysis we observe that they cover the majority of cases. 
These use cases were impossible or required a work-around before the quote RT feature was introduced. For example, for a public reply users would edit the reply, typically by adding a leading `.', so that Twitter treats the tweet as a mention, instead of as a reply.\footnote{\url{https://support.twitter.com/articles/119138}} Forwarding could be achieved by mentioning the recipient and manually copy-pasting the forwarded tweets URL. Adding an opinion required manually shortening the quoted tweet and using the remaining characters for the comment. 
Note that each of the three use cases has the potential to improve political discourse and information spread, in particular as normal replies are not shown to a user's followers. The new feature also helps to put a user's opinion in context whereas a normal reply or mention would lack the context of the original tweet.




From the above analysis, we see that quote RTs are being used partly to replace the `reply' (Public Reply) and `mention' (Forwarding) functionality on Twitter. 
To understand more on how this new feature is used in comparison to other existing forms of interactions (retweet, mention and reply), we (i) looked at the fraction of tweets from different types of interactions over time, (ii) compared users who use these features, (iii) analyzed differences in their follow/friend patterns, and (iv) examined if this feature helps diffuse information.

\subsection{Fraction of tweets} 

\begin{figure}
\centering
\includegraphics[width=0.35\textwidth]{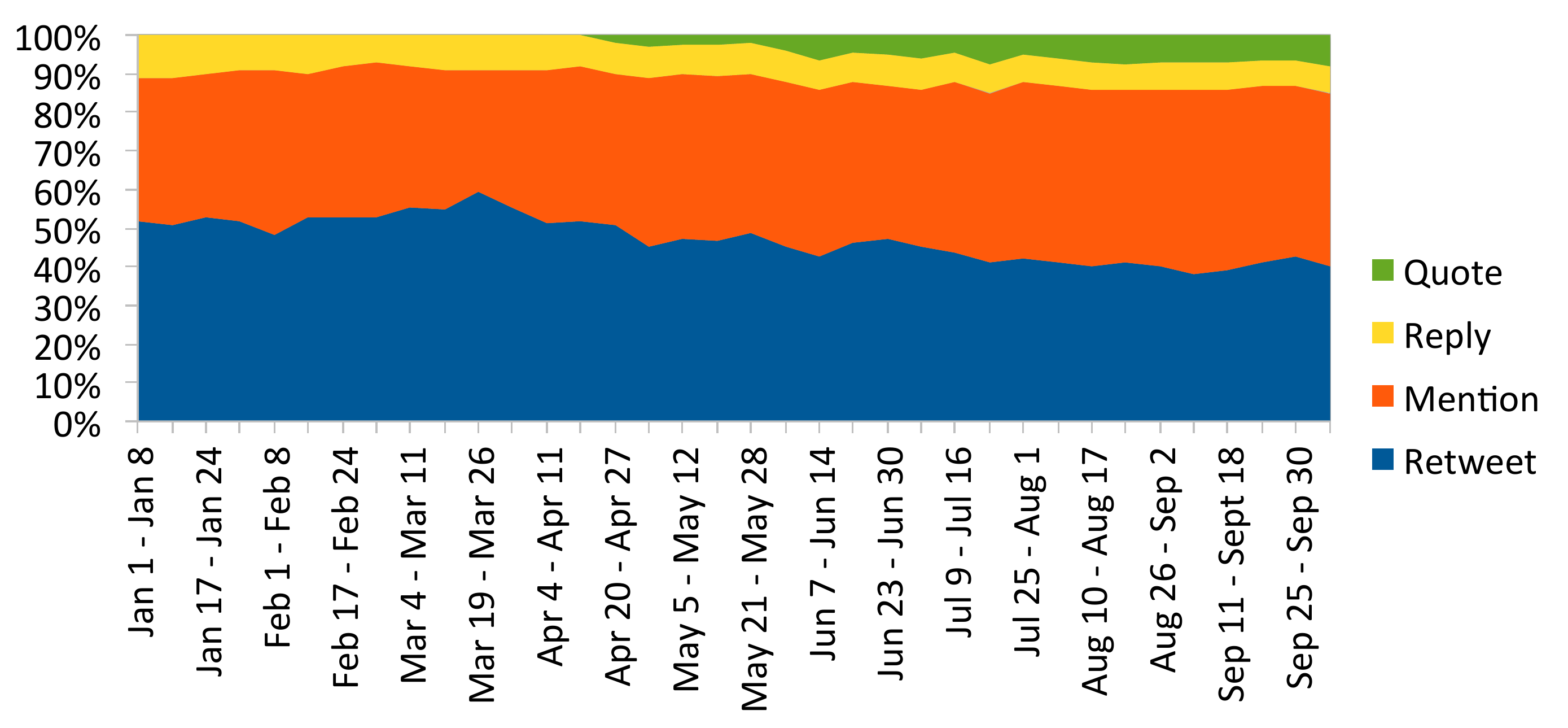}
\caption{Percentage of tweets of various types across time.}
\label{fig:fraction_tweets}
\vspace{-2mm}
\end{figure}

We first looked at the fraction of tweets from the four major forms of interactions on political content on a weekly basis from Jan--Sep 2015. The results are presented in Figure~\ref{fig:fraction_tweets}. It appears from the figure that quote RTs are being used as a replacement for replies and retweets.

\subsection{Users adopting this feature}
\begin{figure}
\centering
\includegraphics[width=0.38\textwidth, height=0.35\textwidth, clip=true, trim=25 50 25 45]{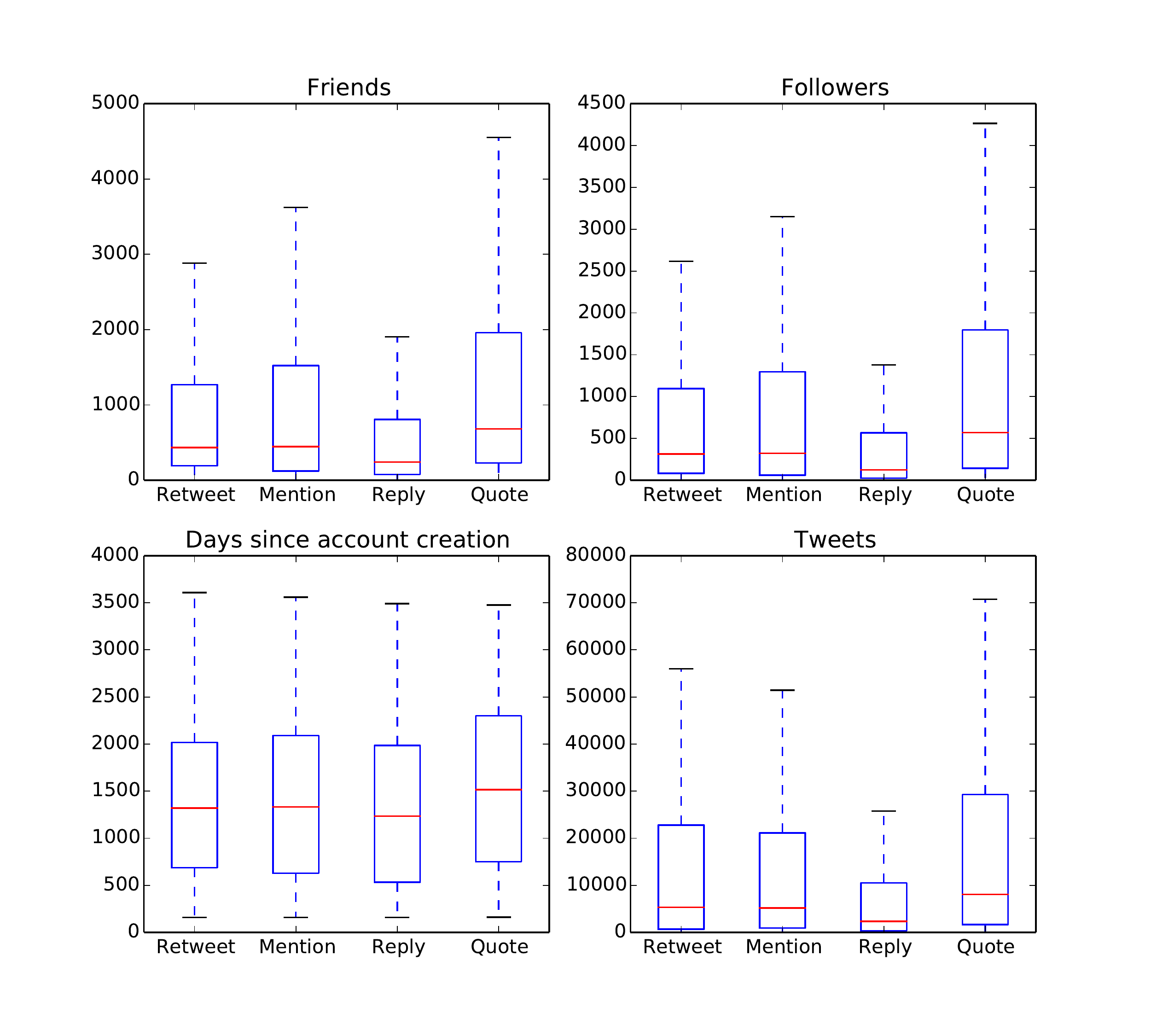}
\caption{Comparison of users who use various types of interactions. Users involved in a political interaction during the month of September 2015 are considered.
The whiskers in the box plots indicate the lowest and highest datum with in 1.5 times the inter quartile range (IQR). 
}
\label{fig:user_features}
\vspace{-2mm}
\end{figure}

For measuring the behavior of users who use various types of interactions, we used only those users who were involved in a political interaction (retweet, mention, reply or quote RT) during September 2015. This choice (of September) was made to reduce the impact of factors related to early adoption (of quote RTs). For each user, we considered four parameters: friends, followers, number of tweets, and days since they created their Twitter account.
The results of this analysis are presented in Figure~\ref{fig:user_features}. 
The results show that quote RTs are being used by users who are more social - more friends, followers, tweets and have been on Twitter longer. The differences in the means are all statistically significant at $p<0.01$.


\subsection{Mutual following}
Liu and Weber~\cite{liu2014twitter} show that if two users follow each other and mention each other, it is most likely to be a tweet containing political alignment. On the other hand, for users who do not follow each other, a mention is most likely going to indicate political disalignment. Given how this mutual following creates a difference in the political interactions, we analyzed how different directed communication patterns are for quotes and compared them with other forms of interaction.
For computing this, we looked at each pair of interacting users (Ex.\ seed user and the user retweeting them) and averaged their mutual following counts for May--Sept 2015.

The results are presented in Table~\ref{tab:bidir}. From Table~\ref{tab:bidir} we can see that quote RTs behave mostly like replies. A higher fraction of mutual following and the lowest fraction of no direction follow indicate that quote RTs are used mostly within settings of political alignment.




\begin{table}
\centering
\begin{tabular}{ | l | l | l | l | l | }
\hline
	& retweet & mention & reply & quote \\ \hline
	S$\leftrightarrow$U & 9.1 & 8.6 & 7.8 & 11.8 \\ \hline
	S$\rightarrow$U & 52.9 & 42 & 63.5 & 60 \\ \hline
	S$\leftarrow$U & 0.6 & 0.9 & 0.6 & 0.4 \\ \hline
	S---U & 37.4 & 48.5 & 28.1 & 27.8 \\ \hline
\end{tabular}
\caption{Bi Directional follow percentage for various types of interactions. `S': seed user, `U': user, S$\rightarrow$U indicates U follows S, S --- U indicates follow in neither direction.}
\label{tab:bidir}
\vspace{-3mm}
\end{table}

\subsection{Does information spread further?}
\label{sec:network}

Next, we examine if quote RTs help diffuse information on the social network. 
To evaluate the information spread, we look at (i) whether users are directly connected and (ii) whether they are part of the same `clique'.
In particular, we computed the fraction of a user's followers who also follow the seed user. If this fraction is high, then the content has left its original clique.

\begin{figure}
\centering
\includegraphics[width=0.23\textwidth,height=0.2\textwidth, clip=true, trim=5 20 50 20]{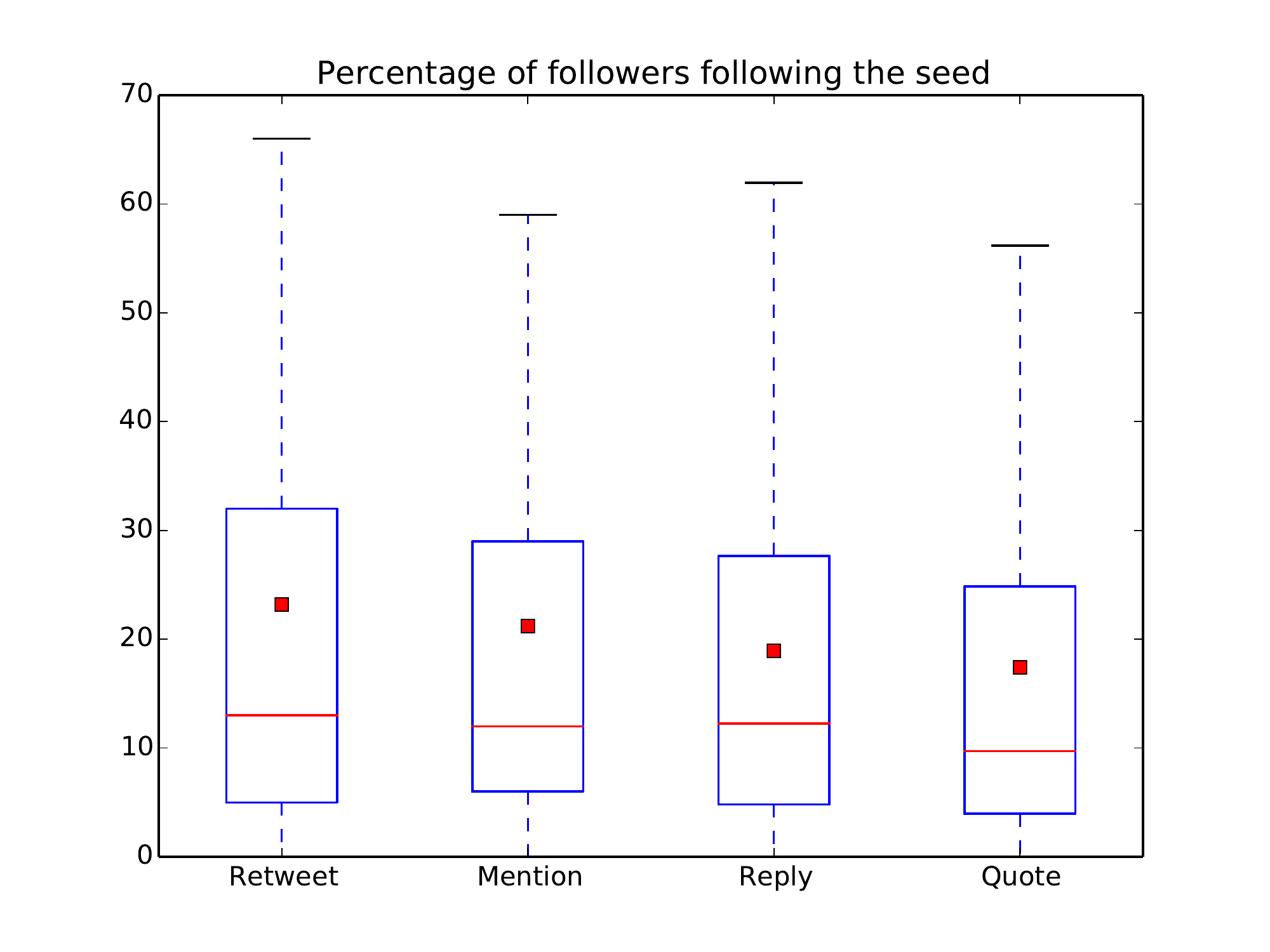}
\caption{Percentage of a user's followers who also follow the seed user. The red line indicates the median and the red dot indicates the mean, whiskers with maximum 1.5 IQR.}
\label{fig:fraction_following}
\vspace{-2mm}
\end{figure}

The results are shown in Figure~\ref{fig:fraction_following}. We observe a higher fraction for replies, indicating that replies are mostly within politically oriented users. On the other hand, the significantly lower fraction for quote RTs indicates that they help to spread political discourse further from its original source. The differences shown in Figure~\ref{fig:fraction_following} are statistically significant at $p<0.01$ (measured using a Welch's $t$-test).

We also measured if there is more interaction with quotes compared to replies, in terms of the average number of retweets and favorites that quote RTs and replies get. These numbers are significantly higher for quote RTs than replies (2.24 vs.\ 0.21 mean retweets, 2.56 vs.\ 0.56 mean favorites). This provides additional support about the increased visibility provided by this feature.



\section{Civility of the Discourse}




In this section, we analyze the civility of the discourse that is being created by quote RTs. We measure two aspects: (i) Agreement and disagreement between the quote and the original tweet, and (ii) insults in quote RTs.

To have a comparative baseline in the existing forms of interactions, we chose to compare only with replies.
We do not compare the quote RTs to mentions or retweets because replies and quote RTs are the most similar in their interactional and conversational aspects. Retweets, for Ex.\ re-broadcast an existing message without creating a conversation. Mentions that are not replies are also not used in a conversation, where replies would be used instead.




%

First, to measure agreement and disagreement, we set up a CrowdFlower\footnote{\url{http://www.crowdflower.com/}} task, and paid human judges for labeling our data. 
%
The task showed two tweets, the original tweet and the quoting/replying tweet. We asked crowd workers to rate the agreement/disagreement between the latter and the former tweet. There were three options: (i) Agreement, (ii) Disagreement, and (iii) No agreement/disagreement.

The results are shown in Table~\ref{tab:agreement}. The Fleiss $\kappa$ for $N=1,695$, 3 users per label is 0.38 for quotes and 0.4 for replies.
The Fleiss $\kappa$ is in the `fair' range according to~\cite{landis1977measurement}. 
The low values of $\kappa$ might be because judging agreement and disagreement with out complete context (just by looking at the tweets) is a hard task.
A higher percentage of agreement, less disagreement and majority of netural tweets for quote RTs shows that political conversations are now more engaging and balanced, when compared to replies, which have a comparitively much larger fraction of disagreement. 

\begin{table}
\centering
\begin{tabular}{ | l | l | l | l | }
\hline
	& agree & disagree & neutral \\ \hline
	quote & 27.6\% & 4.2\% & 68.2\% \\ \hline
	reply & 28.1\% & 28.9\% & 43\% \\ \hline
\end{tabular}
\caption{CrowdFlower agreement/disagreement results}
\label{tab:agreement}
\vspace{-4mm}
\end{table}


Next, we investigated if the new feature has increased hate speech and insults, especially in the context of political discourse. 
To identify a potential source of insults, we used LIWC categories `swear' and `death' to first identify tweets containing swear words. The category death was chosen to include harassment threats such as `kill', `die', etc.
%
Similar to the above task, we used Crowdflower to label if quotes and replies containing swear words were actually meant as an insult or even hate speech as opposed to strong but agreeing language as in ``Yes, f**king exactly right!''. 
The task showed two tweets, the original tweet and the quoting/repling tweet. The crowd workers were asked to label if the quote/reply was meant as an insult to the original user.
66\% of replies were insults, where as 58\% of quotes were insults.
Fleiss kappa 0.52 for quotes and 0.55 for replies (3 users per label, $N=1,650$ in each case).
The reason for lower fraction of insults might be due to the way the feature is designed. On Twitter, replies are `semi-private' between two users and not shown by default on other users' home timelines. On the other hand, quote RTs are regular tweets, which have much more visibility.






\section{Case study}

\begin{figure}[ht] 
  \begin{subfigure}[b]{0.5\linewidth}
    \centering
    \includegraphics[width=0.75\linewidth, clip=true, trim=0 80 0 40]{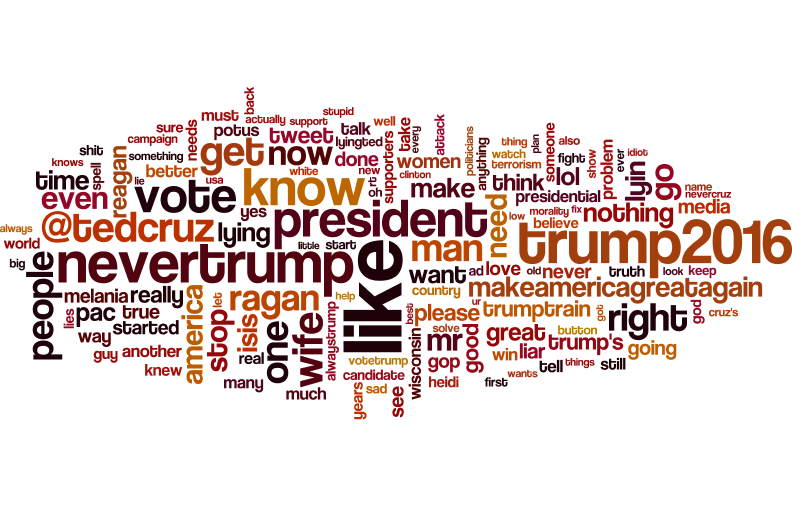} 
    \caption{Donald Trump} 
    \label{fig7:a} 
  \end{subfigure}
  \begin{subfigure}[b]{0.5\linewidth}
    \centering
    \includegraphics[width=0.75\linewidth, clip=true, trim=0 80 0 40]{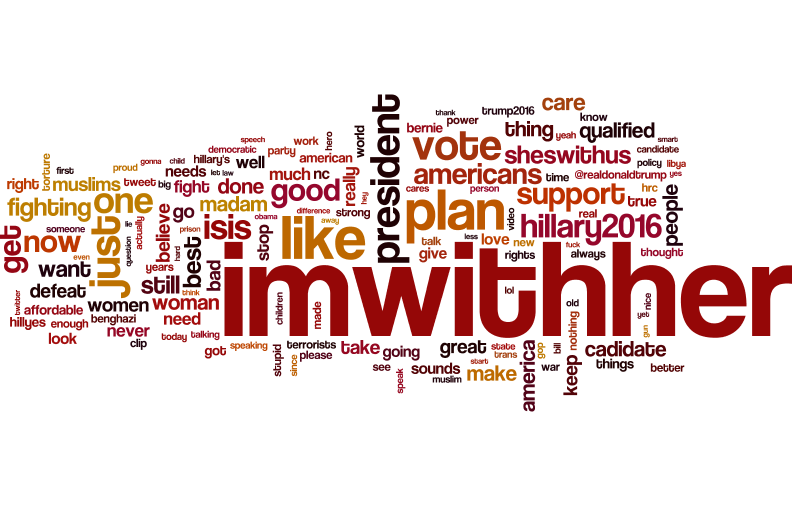} 
    \caption{Hillary Clinton} 
    \label{fig7:b} 
  \end{subfigure} 
  \begin{subfigure}[b]{0.5\linewidth}
    \centering
    \includegraphics[width=0.75\linewidth, clip=true, trim=0 40 0 40]{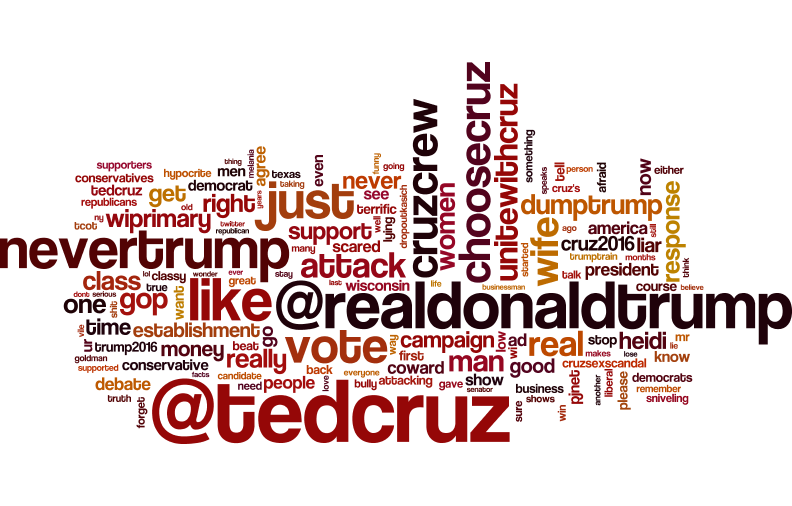} 
    \caption{Ted Cruz} 
    \label{fig7:c} 
  \end{subfigure}
  \begin{subfigure}[b]{0.5\linewidth}
    \centering
    \includegraphics[width=0.75\linewidth, clip=true, trim=0 0 0 20]{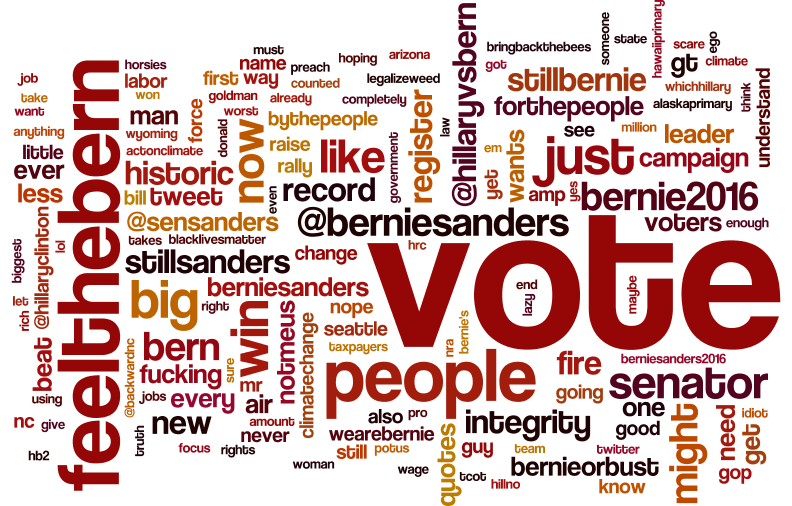} 
    \caption{Bernie Sanders} 
    \label{fig7:d} 
  \end{subfigure} 
  \caption{Wordclouds of Quote RTs from the front runners of the 2016 US Presidential elections.}
  \label{fig:wordclouds} 
\vspace{-2mm}
\end{figure}

To empirically evaluate our hypothesis about the impact of quote RTs on political discourse, we collected quote RTs from the front runners of the US presidential elections. We monitored the Twitter streaming API for quote RTs of the candidates twitter handles for 3 days (Mar 22-24). This way, we obtained $\approx$35,000 quote RTs.
Figure~\ref{fig:wordclouds} shows word clouds (with the candidate names removed) generated from the quote texts for the four candidates. We can observe that quote RTs enable a wide range of actions, including (i) discussion (e.g.\ mentions of other candidates), (ii) agreement (e.g.\ feelthebern or imwithher), (iii) disagreement (e.g.\ nevertrump or dumptrump), etc.

\section{Limitations and Discussion}

Our study provides some of the first insights into how the new quote RT feature on Twitter has been catering to political discourse. The observation that political information diffuses further in the Twitter social network when this mode of interaction is used, goes on to show that its usage might be associated with receding polarization on the platform. While causality cannot be directly inferred, we can still safely conclude that different modalities of social media platforms can play different roles in how ideological echo chambers form, grow, or shrink. Increased civility of political discourse and the greater involvement of socially active Twitter users further show that platform-related changes (e.g., inclusion of a new interactional modality) not only engages different users differently, but also can go to great lengths in allowing social media sites evolve into platforms of more constructive exchange.

There are some limitations to our study. %
First, conversations could be driven by external events and so the quality of discourse might change as elections approach. For example, in our previous (separate, unpublished) work, using the methodology from~\cite{weberetal13asonam} applied to US politics, we observed a saw-tooth pattern of polarization: polarization would build up towards US presidential elections, then sharply drop, then build up to a smaller peak for the mid-term elections, and the drop again before building up towards the next presidential elections. Therefore, we can not exactly pinpoint that the changes after the introduction of this feature might be due to this feature.
%
Second, the usage of this new feature might not have `converged' yet and it could still be evolving. In particular, the usage could still be limited to early adopters with characteristics different from other users. However, Figure~\ref{fig:user_features} seems to indicate that the users of the new feature are largely similar to those of other features. 
%
%
Third, Twitter as a whole continues to evolve \cite{liu2014tweets} and newer, younger users might use the tool differently. Therefore, any study needs to be interpreted within the context of its time. 
%
%
Despite these limitations, we believe studies such as ours are important to better understand how the medium affects the message. Especially with the rise in partisan politics and online trolling, it is timely to study which features can help to break filter bubbles and are less prone to be abused.

\section{Conclusions}

In this paper, we presented the first study of a Twitter's new quote RTs feature in the context of political discourse. 
We first investigated what new means of communication were enabled by this new feature. We found out that, (i) when compared to existing forms of interaction, quote RTs are similar in their usage to reply to an original tweet, more than as a form of retweet. This is evident both from a qualitative inspection and in terms of mutual following patterns; (ii) users adopting this new feature are usually more socially active and have been on Twitter for longer; and (iii) the new feature helps spread political discourse `outwards' of the social network of the users.

We performed crowd labeling to understand the civility of the discourse enabled by the new feature and found that quote RTs enable a more civilized form of communication where people discuss and agree with each other, with far fewer insults being observed.


Overall, this provides evidence that the change to the medium, i.e., Twitter's new feature, has led to a slightly positive change to the message, i.e., a more civil political discourse. Time will tell whether eventually quote RTs will become the new playground for trolls.

\bibliographystyle{abbrv}
\bibliography{bibliography}

\end{document}